\documentclass[prd,aps,twocolumn,superscriptaddress,floatfix,nofootinbib,10pt]{revtex4-1}

\usepackage{setspace} 
\usepackage[utf8]{inputenc}
\usepackage{float}
\usepackage{amsmath,amssymb,amsfonts,bm}
\usepackage{graphicx}
\usepackage{multirow}
\usepackage{booktabs,tabulary}
\usepackage[usenames,dvipsnames]{color}
\usepackage{siunitx}
\usepackage{cancel}
\allowdisplaybreaks
\usepackage[
colorlinks=true,
linkcolor=blue,
breaklinks=true,
urlcolor=blue,
citecolor=blue]{hyperref}
\usepackage{pgf}
\usepackage{lmodern}
\usepackage{import}
\usepackage{bbm}

\usepackage{orcidlink}

\definecolor{green}{rgb}{0,0.6,0}

\newcommand{\Omnes}{Omn\`{e}s }
\newcommand{\Disc}{\textrm{Disc}}
\newcommand{\Mcal}{\ensuremath{\mathcal{M}}} 
\newcommand{\Acal}{\ensuremath{\mathcal{A}}} %
\newcommand{\Bcal}{\ensuremath{\mathcal{B}}} %
\newcommand{\diff}{\mathrm{d}}

\newcommand{\GeV}{\,\text{GeV}}
\newcommand{\MeV}{\,\text{MeV}}

\newcommand{\Order}{\mathcal{O}}

\newcommand{\bonn}{\affiliation{Helmholtz-Institut f\"ur Strahlen- und Kernphysik and Bethe Center for Theoretical Physics,\\ Universit\"at Bonn, 53115 Bonn, Germany}}

\newcommand{\ucm}{\affiliation{Departamento de Física Teórica and IPARCOS, Universidad Complutense de Madrid, 28040 Madrid, Spain}}

\newcommand{\camp}{\affiliation{Departamento de Raios Cósmicos e Cronologia, Universidade Estadual de Campinas 13083-860, Campinas, Brazil}}

\newcommand{\fzj}{\affiliation{Institute for Advanced Simulation (IAS-4), Forschungszentrum J\"ulich, 52425 J\"ulich, Germany}}

\newcommand{\usi}{\affiliation{Center of Particle Physics (CPPS), Theoretische Physik 1, Universit\"at Siegen, 57068 Siegen, Germany}}

\begin{document}

\title{Understanding large localized CP violation in \boldmath{$B^\pm\to K^\pm\pi^+\pi^-$}\\ using dispersive methods}

\begin{abstract}
\noindent
We utilize the universality of pion--pion ($\pi\pi$) final-state interactions at small invariant masses to understand the enhanced localized CP violation in $B^\pm\to K^\pm\pi^+\pi^-$, using a dispersive approach. From a fit to the integrated CP-asymmetry data, we successfully predict the Dalitz-plot kinematic distribution of the asymmetry in the low-energy $\pi\pi$ region, including the large localized CP violation recently observed by LHCb. An essential role is played by the contributions of isospin 2. This formalism, whose parameters have a physical meaning, can be adapted straightforwardly to other systems with CP violation enhanced by final-state interactions.
\end{abstract}

\author{L. A. Heuser\orcidlink{0009-0006-0410-6269}$^a$}\email{heuser@hiskp.uni-bonn.de}
\renewcommand{\thefootnote}{\alph{footnote}}\footnotetext{These authors contributed equally to this work.}
\bonn
\author{A. Reyes-Torrecilla\orcidlink{0009-0006-8336-2323}$^a$}\email{albrey01@ucm.es}
\ucm
\author{C. Hanhart\orcidlink{0000-0002-3509-2473}}
\fzj
\author{B.~Kubis\orcidlink{0000-0002-1541-6581}}
\bonn
\author{P. C. Magalh\~aes\orcidlink{0000-0003-3641-8110}}
\camp
\author{T. Mannel\orcidlink{0009-0004-9310-5676}}
\usi
\author{J. R. Pel\'aez\orcidlink{0000-0003-0737-4681}}
\ucm\bonn

\maketitle

{\it Introduction}---Non-leptonic decays of heavy mesons are notoriously difficult to describe theoretically.  Although QCD factorization methods exist for two-body decays, three-body decays remain a challenge~\cite{Krankl:2015fha}. In the Standard Model, CP violation (CPV) is induced by 
an interplay of the weak phase from the Cabibbo--Kobayashi--Maskawa (CKM) matrix~\cite{Cabibbo:1963yz,Kobayashi:1973fv} and the strong phases of hadronic matrix elements.  The importance of three-body decays is that strong phases depend
on two kinematic variables, resulting in a complex structure in the Dalitz-plot distribution of CP asymmetries, thus posing a very challenging test for our understanding of final-state interactions (FSI) in CPV.  The experimental breakthrough came when LHCb~\cite{Aaij:2013sfa,Aaij:2013bla,Aaij:2014iva} measured these distributions accurately for $B^\pm\to h_1^\pm h_2^+ h_2^-$ decays, where $h_n^\pm=\pi^\pm,K^\pm$.

While the inclusive CP asymmetries are only a few percent~\cite{Aaij:2014iva}, LHCb observed, and reaffirmed in their recent $5.9\,\text{fb}^{-1}$  
analysis~\cite{LHCb:2022fpg}, one order of magnitude larger asymmetries in localized Dalitz-plot regions. Through amplitude analyses with simple resonance models~\cite{Aaij:2019hzr,Aaij:2019jaq,Aaij:2019qps}, LHCb has attributed these large localized asymmetries to hadronic FSI, particularly in the $S$~wave. For instance, they claim that the $\pi\pi\rightarrow K\bar K$ $S$-wave rescattering contribution to $B^\pm\rightarrow \pi^\pm K^+K^-$ has {\em ``the largest CP asymmetry reported to date for a single amplitude of $(-66\pm 4 \pm 2\%)$''}~\cite{Aaij:2019qps}.  The dominant role of $S$-wave $\pi\pi \leftrightarrow K\bar K$ also naturally explains~\cite{Bhattacharya:2013cvn,Bediaga:2013ela,AlvarengaNogueira:2015wpj,Garrote:2022uub} the opposite integrated CP asymmetries in $B^\pm \to \pi^\pm\pi^+\pi^-$ and $B^\pm\to \pi^\pm K^+ K^-$ in the 1 to $1.5 \GeV$ invariant-mass region of the pair $h_2^+ h_2^-$.

The rich CPV Dalitz-plot structure of $B$-meson decays cannot be reproduced by a leading-order calculation within the QCD factorization approach, even in terms of pseudo-two-particle processes~\cite{Mannel:2020abt}.  Still, an approach that treats the hadronic part of the decay systematically is of utmost importance in any CPV analysis.  In this Letter, we bring forth an important step in this direction.

To illustrate the method in its simplest form, we focus on $B^\pm\to K^\pm\pi^+\pi^-$, as the kaon spares us from symmetrization complications and because the observed large localized CPV lies at the $m_{\pi\pi}\leq 1.5\GeV$ invariant-mass region (cf.\ Fig.~3 in Ref.~\cite{LHCb:2022fpg}), which is particularly intriguing. In existing $B^\pm\to h_1^\pm h_2^+ h_2^-$ amplitude analyses~\cite{Aaij:2019qps,Aaij:2019jaq}, resonances are commonly included as Breit--Wigner (BW) parametrizations or variants thereof within isobar formalisms; however, this approach is in general model- and channel-dependent, and if there is more than one resonance in a partial wave or complex couplings are used, it violates unitarity.  In addition, it does not allow for a proper inclusion of chiral symmetry constraints, nor a description of the isoscalar scalar $\pi\pi$ scattering data and the non-BW-like $f_0(500)$ and $f_0(980)$ resonances that appear therein~\cite{Meissner:2003pd}.  Moreover, non-resonant partial waves are usually ignored.  By construction, our method is consistent with unitarity and incorporates the high-accuracy phase-shift information available for the two-pion system~\cite{Ananthanarayan:2000ht,Colangelo:2001df,Garcia-Martin:2011iqs,Caprini:2011ky,Pelaez:2024uav}.
As a first step, we develop our method within the elastic $\pi\pi$ scattering region, in practice $m_{\pi\pi}<2M_K\sim 1\GeV$, since it contains most of the large localized CP asymmetry in $B^\pm\to K^\pm\pi^+\pi^-$, which appears with both signs in this mass regime. 

Besides FSI universality, we make three assumptions: 

1. The matrix elements of the weak Hamiltonian are treated as short-distance sources, constructed in the spirit of~\cite{Meissner:2000bc}, for the
outgoing hadrons, which are ``dressed'' subsequently with their final-state interactions.

2.  For small $m_{\pi\pi}$, the pion pair has a large recoil against the $BK$ system, of the order of the $B$-meson mass, $M_B$. 
Hence, the $ K\pi$ invariant mass is $\Order(M_B)$, and the kaon final-state rescattering (crossing) is expected to be negligible since factorization becomes exact in the limit of an infinitely heavy decaying particle with a finite invariant mass of the pion pair. This can be tested in the future using a Khuri--Treiman three-body formalism~\cite{Khuri:1960zz}.

3. Left-hand cuts, e.g., from  $B\to D^*\bar D_s$, followed by $D^*\to D\pi$, with the remaining $D\bar D_s$ system transitioning to $\pi K$, can be approximated as providing (CP-even) constant imaginary parts and, as such, subsumed in the source terms.

In such a picture, all relevant energy dependencies and singularities are driven by the FSI of the outgoing pion pair, of long-distance nature and thus universal. 

{\it Notation}---The $B^\pm(p_B)\to K^\pm(p_K) \pi^+(p_+)\pi^-(p_-)$ decay is characterized by the  $s=(p_+ + p_-)^2\equiv m_{\pi\pi}^2$ and  $t=(p_K+p_\mp)^2=(p_B-p_\pm)^2\equiv m_{K\pi}^2$
Mandelstam variables.
In the following, we work with partial-wave amplitudes of given angular momentum and isospin $I$.  For $m_{\pi\pi}<1\GeV$, those of relevance are the isoscalar $S0$~wave, featuring the $f_0(500)$ and $f_0(980)$ resonances, the isovector $P$~wave with the prominent $\rho(770)$, and the isospin-2 $S2$~wave, free of resonances. 
We checked that $D$~waves have a negligible effect in this range. 
Since the $f_0(980)$ lies almost at the $K\bar K$ threshold and couples strongly to two kaons, it is convenient to use a $\pi\pi$--$K\bar K$ coupled-channel formalism, even in the elastic region, and split the $S0$~wave according to
$\Acal^{\pm}_{S0}=\Acal^\pm_{S0n}+\Acal^\pm_{S0s}$, where the two parts are connected to the non-strange- and strange-quark operator structures of the source. The decay amplitude has the compact form
\begin{equation}
\Acal^\pm(s,t)=\sum_{i} f_i(s,t)\Acal_i^\pm(s) 
\end{equation} 
with $i\in \{S0n,S0s,S2,P\}$.
The dependence on the angle $\theta$ between the $K^\pm$ and the $\pi^\mp$
is carried by $f_i(s,t)=1$ for $i=S0n,S0s,S2$ and $f_P(s,t)=t-u=g(s)z(s,t)$, where $z(s,t)=\cos\theta$. Here $g(s)=-\sigma_\pi(s) \lambda^{1/2}(s,M^2_K,M_B^2)$, $\sigma_h(s)=\sqrt{1-4M_h^2/s}$, and $\lambda(x,y,z)=x^2+y^2+z^2-2xy-2yz-2xz$. 

{\it Universal final-state interaction}---The starting point for the dispersive analysis is the discontinuity relation for the production partial-wave amplitudes along the pion-pair cut at $s>4M_\pi^2$.  For $i=S2,P$ this reads
\begin{equation} 
  \Disc \, \Acal^{\pm}_i(s)
   = 2i \rho_\pi(s)\Mcal_i(s)^* 
   \Acal^{\pm}_i(s),
   \label{resonances:eq:discAmain}
\end{equation}
where $\rho_h(s)=\sigma_h(s)/16\pi$. Note the appearance of the $\pi\pi$ scattering
partial-wave amplitude $\Mcal_i(s)$, which, being CP-invariant,
carries no $\pm$ superscript. For the elastic case,
$\Mcal_i(s)=e^{i\delta_i(s)}\sin\delta_i(s)/{\rho_\pi(s)}$, where $\delta_i$ is its phase shift. This leads to a closed-form solution~\cite{Omnes:1958hv}:
\begin{equation}
\Acal^{\pm}_i(s)=P_i(s)\Omega_i(s) \bar \Acal_i^{\pm},
\end{equation}
which describes the low-energy universal pion-pair interactions, contained in the \Omnes function for the elastic case: 
\begin{equation}
\Omega_i(s)=\exp\left\{\frac{s}{\pi}\int_{4M_\pi^2}^\infty
\diff s' \frac{\delta_i(s')}{s'(s'-s-i\epsilon)}\right\}, \ \Omega_i(0)=1.
\end{equation}
The constants $\bar\Acal_i^{\pm}$ parametrize the source, and $P_i(s)$, with $P_i(0)=1$, is a function free of right-hand cuts up to inelastic thresholds, to be parametrized by a polynomial.  This form automatically satisfies Watson's theorem~\cite{Watson:1954uc} (cf.\ also Refs.~\cite{Wolfenstein:1990ks,Suzuki:1999uc,Bediaga:2013ela,Cheng:2016shb}), stating that in the elastic case, the phases of the production and scattering partial-wave amplitudes coincide.
The $P_i(s)$ are often well approximated by 1 (cf.\ Ref.~\cite{Daub:2015xja} for $B\to J/\psi \pi^+\pi^-$), or, at most, as linear polynomials, whose slopes are free parameters.  While the \Omnes function captures all the universal elastic hadron--hadron
interactions, the $P_i(s)$ are reaction-specific and absorb the uncertainty in the high-energy extension of the phases and, to some extent, crossed-channel effects~\cite{Kubis:2015sga}. 

For the $S0$~wave, Eq.~\eqref{resonances:eq:discAmain} in coupled channels just has one more term accounting for $K\bar K\to\pi\pi$ rescattering:
\begin{align}
    \Disc \Acal^{\pm}_{S0n}&=2i\left(\Mcal_{11}\rho_\pi \Acal^{\pm}_{S0n}+\Mcal_{12}^*\rho_K \Bcal^{\pm}_{S0n}\right),\notag \\
    \Disc \Acal^{\pm}_{S0s}&=2i\left(\Mcal_{11}\rho_\pi \Acal^{\pm}_{S0s}+\Mcal_{12}^*\rho_K \Bcal^{\pm}_{S0s}\right)\label{eq:coupled_channel},
\end{align}
where $\Bcal^{\pm}_i$ is the corresponding $K\bar K$ production amplitude and, for brevity, we have suppressed the $s$ dependence. The $\Mcal_{ij}(s)$ are the scalar scattering partial-wave amplitude elements between channels $1=\pi\pi$ and $2=K\bar K$. They are parametrized in terms of the $\pi\pi\to\pi\pi,K\bar K$ phase shifts together with the elasticity $\eta(s)$~\cite{Buettiker:2003pp,Pelaez:2018qny,Pelaez:2020gnd}.\footnote{The coupled-channel $I=0$ $S$-wave formalism has previously been applied in the context of CPV for $D$-meson two-body decays~\cite{Pich:2023kim}.} 
There are no closed-form solutions of Eq.~\eqref{eq:coupled_channel} (and the corresponding discontinuities for $\Bcal^{\pm}_{S0n}$, $\Bcal^{\pm}_{S0s}$), but a matrix solution $\Omega_{ij}$ can be computed numerically from an integral equation, with the conditions $\Omega_{11}(0)=1$, $\Omega_{12}(0)=0$~\cite{Donoghue:1990xh,Moussallam:1999aq,Descotes-Genon:2000byu,Hoferichter:2012wf,Daub:2012mu,Celis:2013xja,Daub:2015xja,Winkler:2018qyg,Ropertz:2018stk}.  Here we use the results of Ref.~\cite{Ropertz:2018stk}. 
The linear combinations corresponding to non-strange and strange sources are then given as
\begin{align}
    \Omega_{S0n}&\equiv\Omega_{11} +\frac{1}{2} \Omega_{12}, & 
    \Omega_{S0s}&\equiv \Omega_{12} .
\end{align}
We allow for a common polynomial $ P_{S0n}(s)\equiv P_{S0s}(s)$ multiplying the complete $S0$~wave.
The complete $B^\pm\to K^\pm \pi^+\pi^-$ amplitude parametrization reads
\begin{equation}
\Acal^\pm(s,t) =\sum_{i} f_i(s,t)P_i(s)\Omega_i(s)\bar\Acal^\pm_{i}  \,,
\end{equation} 

\begin{figure}
\includegraphics[width=\linewidth]{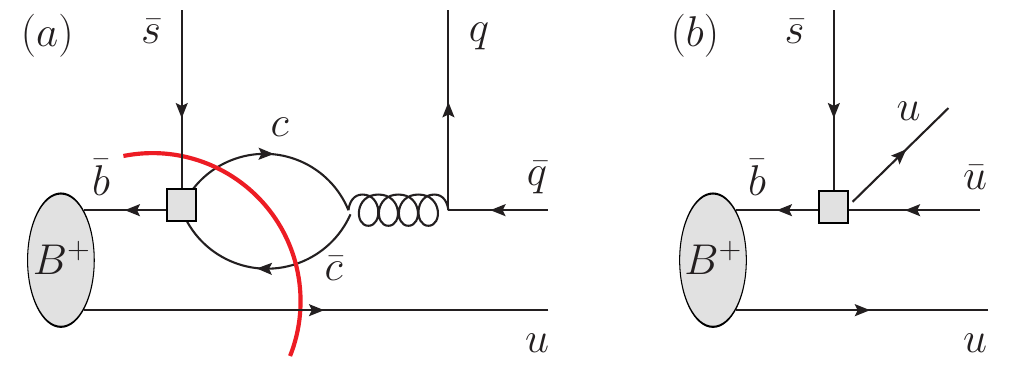}
\caption{Typical topologies for quark-level decays of $B^+\to K^+\pi^+\pi^-$ (a) with and (b) without charm loop. The gray box represents the flavor-changing weak decay process, which contains the CKM elements and CPV phase.
For $q=s$, the operator in diagram (a) provides a $\bar ss$ source,
for $q=u$ or $d$ a $\bar uu$ or $\bar dd$ source.
The red line indicates the cut that generates an imaginary part.
Diagram (b) provides a $\bar uu$ source.
}
\label{fig:ccbar}
\end{figure}

{\it Source terms}---Besides the polynomial slope parameters, the constants $\bar\Acal^\pm_{i}$ remain as free parameters.  They specify the sources and their evaluation involves the two matrix elements $M_{u\bar{u}}$ and $M_{c\bar{c}}$, where $M_{q\bar{q}} = \langle K^+ \pi^+ \pi^- |(\bar{b}q)(\bar{q}s) | B^+ \rangle$ in the limit $m_{\pi\pi}\to0$, plus the CP conjugates for the $B^-$ decay.
Here $(\bar{q} q')$ is the usual weak-interaction current, 
omitting Dirac structures for simplicity.
The relevant effective weak Hamiltonian reads
\begin{equation} \label{Heff}
H_{\rm eff} = \frac{G_F}{\sqrt{2}} \Big(|V_{cb}^* V_{cs}| (\bar{b}c) (\bar{c}s) 
+ e^{i \gamma} |V_{ub}^* V_{us}| (\bar{b}u) (\bar{u}s) 
\Big).
\end{equation}
In the standard parametrization~\cite{Chau:1984fp} of the CKM matrix, $V_{ub}$ carries the weak CPV phase $\gamma$, while $V_{cb}$, $V_{cs}$, and $V_{us}$ are real.
In terms of the Wolfenstein parametrization~\cite{Wolfenstein:1983yz}, $M_{u \bar{u}}$ is suppressed relative to $M_{c \bar{c}}$ by a factor $\lambda^2 \sim 0.04$.  
However, this suppression is partially compensated for by the loop needed to annihilate the $c\bar c$ pair, cf.\ Fig.~\ref{fig:ccbar}(a), whereas the $u\bar u$ pair can be part of the light-meson pair in the final state, cf.\ Fig.~\ref{fig:ccbar}(b).
The $c\bar c$ loop hadronizes as charmed-meson loops, e.g., $\bar{D}^{(*)} D_s^{(*)}$; this mechanism generates CP-even phases (cf.\ Refs.~\cite{Bediaga:2020ztp,Mannel:2020abt}), indicated by the $c\bar{c}$ cut (red line) in  Fig.~\ref{fig:ccbar}(a), beyond those coming from the \Omnes functions.  
These phases are related to the charm-mass short-distance scale and will therefore lead to strongly varying effects near the charm thresholds. 
In the kinematic regions we consider, we do not expect strong energy dependence, and thus the phases can be subsumed in the $\bar\Acal^\pm_{i}$ sources. 
On the other hand, we assume the imaginary parts of the light-meson pair production from $M_{u\bar u}$ before FSI to be negligible, as similar quark-level loops are severely CKM-suppressed, and we neglect residual rescattering with the outgoing kaon. 
The latter could be tackled in the future with more complicated three-body dispersive equations~\cite{Khuri:1960zz,Niecknig:2017ylb,Stamen:2022eda,Kou:2023kvp}. 
As a result, we parametrize the source term as
\begin{equation}
\bar\Acal^\pm_{i} = \hat A_{i}  + e^{\pm i\gamma}\hat B_{i}= a_{i} + i c_{i}  \pm i b_{i} \, ,  \label{eq:amp-parameters}
\end{equation}
where the $\hat A_{i}$ contain a CP-even constant imaginary part due to charm loops, while the $\hat B_{i}$ are real.  The parameters
$a_{i} = {\rm Re} \hat A_{i} + \hat B_{i}  \cos\gamma$,   
$c_{i} = {\rm Im} \hat A_{i}$, and
$b_{i} = \hat B_{i} \sin\gamma$ will be determined by the fit.

Since our source parameters have a physical interpretation, their flavor structure allows us to reduce their number.  First,  $\Acal^\pm_{S0s}$ refers to an isoscalar scalar pion pair emerging from an $\bar ss$ source
through $K\bar{K} \to \pi \pi$ rescattering.
However, the CPV phase $\gamma$ is attached to $M_{u\bar u}$. Accordingly, we set the parameter $b_{S0s}$ to zero.
Moreover,  the $c\bar c$ pair cannot produce a pion pair of isospin 2, and since in our approximation only those matrix elements have a CP-even imaginary part, $c_{S2}$ has to vanish.  Thus, there are three source parameters each for non-strange isoscalar scalar and isovector vector partial waves, and two each for the strange isoscalar scalar and the isotensor scalar ones.  Altogether, there are ten real parameters to parametrize the two CP-conjugated production vertices of four partial waves, plus three real slope parameters in the $P_i$ polynomials. 

The $\pi\pi$ $P$-wave spectrum can be significantly affected by $\rho$--$\omega$ interference in reactions that show strong production of the isoscalar $\omega(782)$.  The smallness of the violation of isospin symmetry is compensated for by an enhancement of $\Order(M_\omega/\Gamma_\omega)\approx 90$ through the narrow $\omega$ propagator.  The strength of $\rho$--$\omega$ mixing can be gleaned directly from the electromagnetic pion form factor if the production strength of the $\omega$ relative to the $\rho$ is known from the flavor structure of the source~\cite{Daub:2015xja}.  As illustrated in Fig.~\ref{fig:ccbar}, diagram $(b)$ provides a $\bar uu$ source, while diagram $(a)$ generates either $\bar uu+\bar dd$ or $\bar ss$.  However, since the $\bar{q}q$ pair from the hard gluon cannot be a colorless source for the pion pair, one of those quarks must combine into the charged kaon.  This prevents the production of a $\bar{d}d$ pair from the gluon.  The resulting pure $\bar uu$ source favors the $\omega$ by a factor of 3 compared to electromagnetic production. We incorporate it by multiplying $\Omega_P$ by $1 + 3 \epsilon_{\rho\omega}s/(M_\omega^2-s-i M_\omega \Gamma_\omega)$ with the known $\rho$--$\omega$ mixing strength $\epsilon_{\rho\omega}$~\cite{Holz:2022hwz,Colangelo:2022prz,Dias:2024zfh}. 
To simplify expressions below, we do not show the $\omega$ contribution individually, but understand it as part of the $\pi\pi$ $P$~wave.

All in all, we arrive at the following expression for the $B^\pm\to K^\pm\pi^+\pi^-$ differential decay count rates 
\begin{align}\nonumber
 &  \Gamma^\pm(s,z)\equiv \frac{\diff^2\Gamma^\pm}{\diff\sqrt{s}\,\diff z}(s,z) =
  -g(s)\sqrt{s}\sum_{i,j}f_if_j P_i P_j \\ \nonumber
   & \times\Big\{ \mbox{Re}\left(\Omega_i\Omega_j^*\right)
 (a_{i}a_{j}+b_{i}b_{j}
 +c_{i}c_{j}
 ) 
+\mbox{Im}\left(\Omega_i\Omega_j^*\right)
 2 a_{i}c_{j}
 \\ 
 & \qquad 
  \pm 2 \Big[\mbox{Re}\left(\Omega_i\Omega_j^*\right)
 c_{i}b_{j}
  + \mbox{Im}\left(\Omega_i\Omega_j^*\right)
 a_{i}b_{j}\Big] \Big\} . \label{eq:amplitude}
\end{align}
Our main result is the CPV difference $\Delta \Gamma_{{\rm CP}}(s,z) \equiv\Gamma^-(s,z)-\Gamma^+(s,z)$, which in the elastic regime reads
\begin{align}\nonumber
& \Delta \Gamma_{{\rm CP}}(s,z) 
= 4g(s)\sqrt{s}\sum_{i,j}f_if_j
P_i P_j |\Omega_i||\Omega_j|
\\
& 
\quad\times 
\Big[ a_{i}b_{j}\sin(\delta_i-\delta_j)+c_{i}b_{j}\cos(\delta_i-\delta_j)\Big].
\label{eq:mainresult}
\end{align}
It is so simple because, in this regime, the production amplitude phase is just
the $\pi\pi$ scattering phase shift $\delta_i$. 
Moreover, the $\Omega_{S0n}$ and $\Omega_{S0s}$ phases are equal. 
The size of $\Delta\Gamma_\text{CP}(s,z)$ is determined by comparing to the sum of differential count rates, i.e., $ \Sigma \Gamma(s,z) \equiv \Gamma^-(s,z)+\Gamma^+(s,z)$,  obviously CP-even.

Our formalism provides several advantages compared to earlier approaches; see, e.g., Refs.~\cite{Simma:1991ct,Fajfer:2004cx,Furman:2005xp,El-Bennich:2006rcn,El-Bennich:2009gqk,Cheng:2016shb,Boito:2017jav,Cheng:2020ipp} (see also the recent reviews~\cite{Bediaga:2020qxg,Chen:2021ftn}). 
Most of these papers start from a factorization of the hadronic matrix elements into a heavy-to-light transition form factor and a light-meson form factor in the timelike region. The latter contains the soft rescattering,  
which, lacking the latest developments in dispersive $\pi\pi$ scattering, were modeled at best by including some $S0$-, $P$-, and $D0$-wave resonances via phenomenological models: isobars, BW formulas with some background, $K$-matrix formalisms, etc. However, non-resonant $\pi\pi$ or $K\pi$ partial waves with $I>1$ are ignored. The \Omnes method was only implemented for the $K\pi$ $S_{I=1/2}$ and $P$~waves~\cite{El-Bennich:2009gqk} for CPV in $B\to K\pi\pi$ decays, omitting the non-resonant $I=3/2$ wave, and never for $\pi\pi$ FSI. Few of those works described projected data~\cite{Furman:2005xp,El-Bennich:2006rcn, El-Bennich:2009gqk}, and none attempted the description of Dalitz-plot regions in detail, as done below. 
Moreover, within QCD factorization, Wilson coefficients contain imaginary parts from charm loops. To this end, a factorization approach can model our source terms, for which we employ a simple parametrization that has a clear physical interpretation.

\begin{figure}
\includegraphics[width=\linewidth]{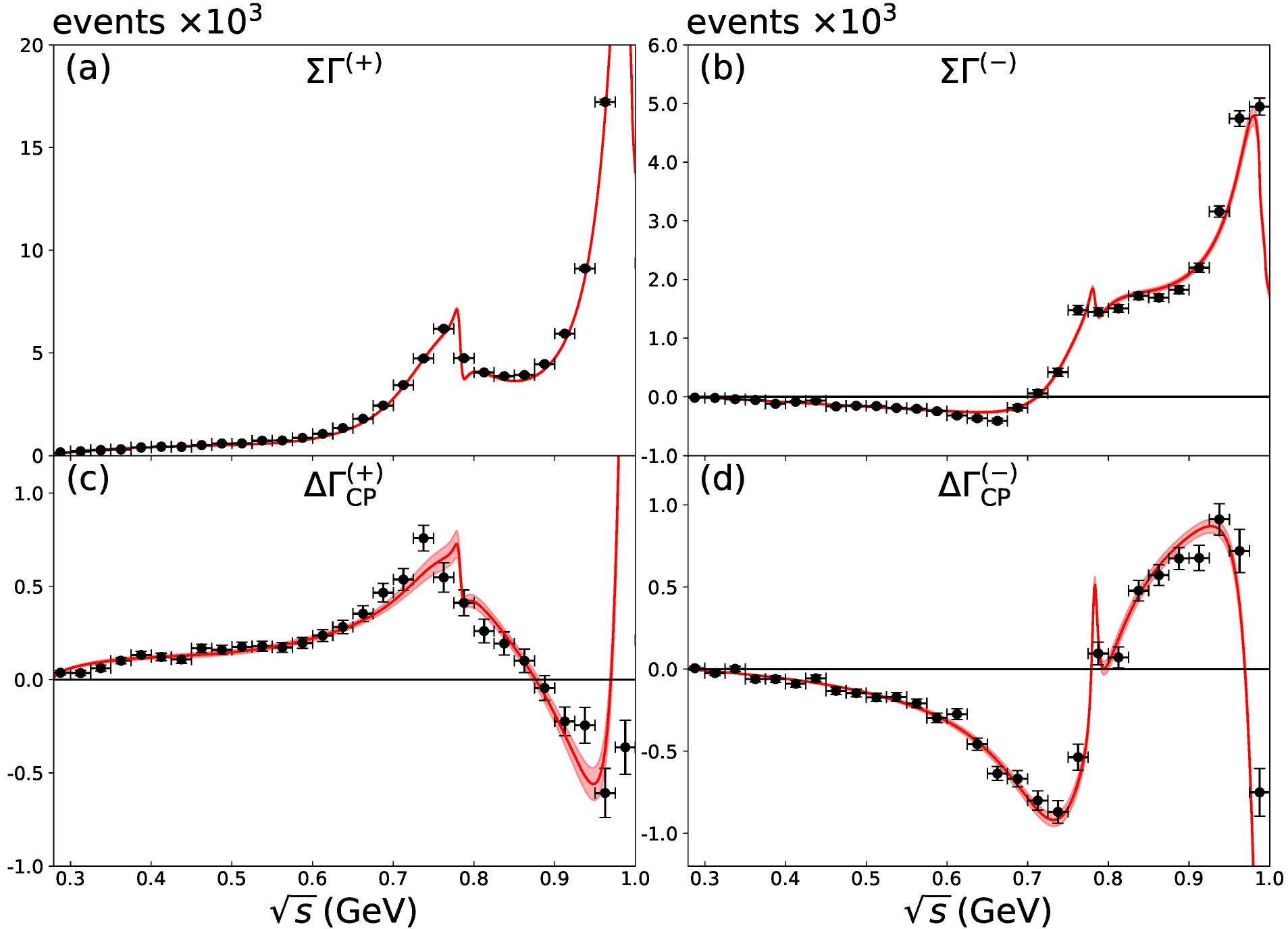}
\caption{Fit to the projected event distributions defined in Eq.~\eqref{eq:observables}, versus LHCb data~\cite{LHCb:2022fpg} (supplemental material). We show
(a) the angle-symmetric sum of yields, 
(b) the angle-asymmetric sum of yields, 
(c) the angle-symmetric CP-violating difference of yields, and 
(d) the angle-asymmetric CP-violating difference of yields.
}
\label{fig:DeltaK}
\end{figure}

Very recently, LHCb~\cite{LHCb:2022fpg} (supplemental material) has provided high-statistics data with uncertainties on $\Gamma^\pm(s,z)$, integrated or ``projected'' separately over the forward ($z=\cos\theta>0$) and backward ($z<0$) regions, to be denoted by $>$ and $<$ superscripts, respectively. In our formalism, the nontrivial $z$ dependence appears linearly in the amplitude through $f_P(s,t)$. 
We can take advantage of these projections by defining
\begin{align}
\Delta \Gamma^{(\pm)}_{{\rm CP}}(s) &=
\Delta \Gamma^>_{{\rm CP}}(s)\pm\Delta \Gamma^<_{{\rm CP}}(s),\nonumber\\ 
\Sigma \Gamma^{(\pm)}(s) &=
\Sigma \Gamma^>(s)\pm\Sigma\Gamma^<(s) ,
\label{eq:observables}
\end{align}
to study differential decay rates not just in CP-symmetric and -antisymmetric
combinations, but simultaneously in forward--backward symmetric or antisymmetric ones. 
The above definitions are useful for disentangling different partial-wave contributions.  Hence, in Fig.~\ref{fig:DeltaK} we show the data following these combinations. 

\begin{figure}[t]
\includegraphics*[trim=2.5cm 0.4cm 2.5cm 11.7cm, clip, width=8.9cm]{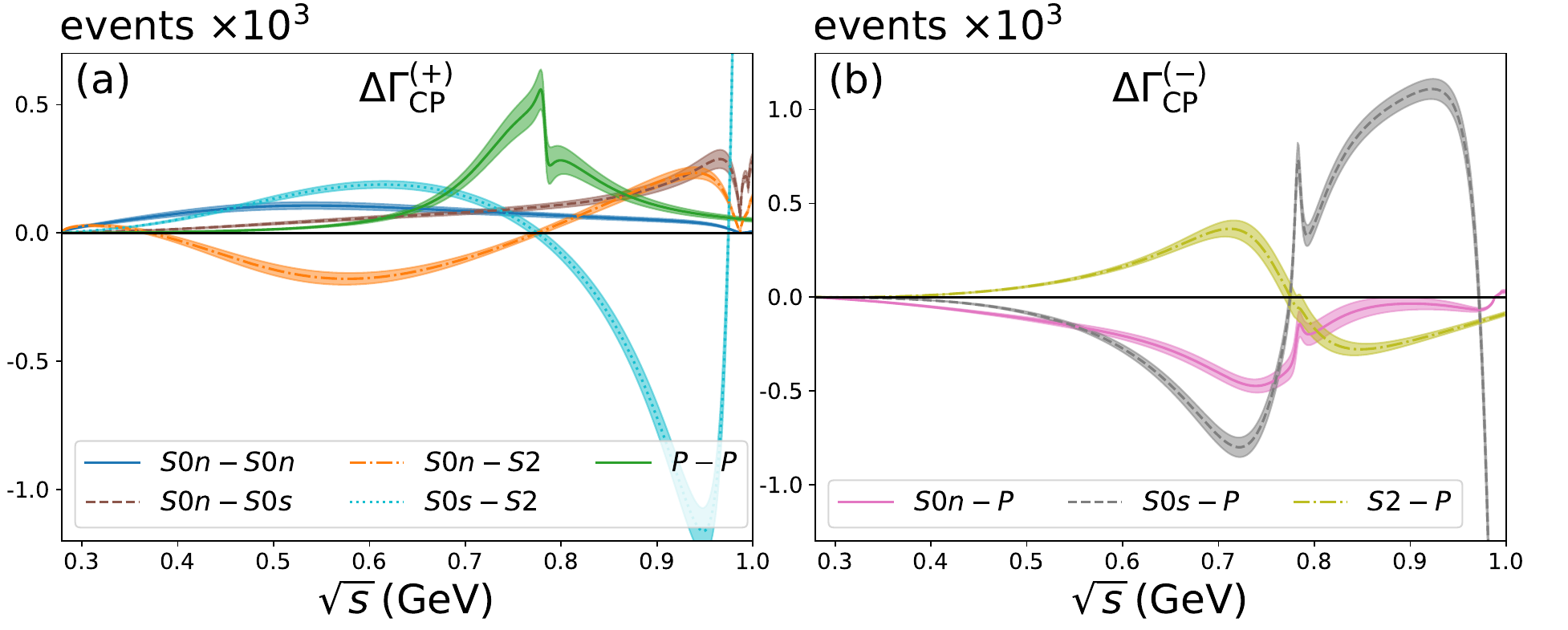}
\caption{
The most relevant contributions to the projected 
(a)~$\Delta\Gamma_\text{CP}^{(+)}(s)$ and
(b) $\Delta\Gamma_\text{CP}^{(-)}(s)$ distributions.
 Note the relevance of the $S2$~wave.}
\label{fig:ContributionsImportant}
\end{figure}

{\it Fit results}---We average our results over the same $25 \MeV$-wide bins in which data were provided.  Since the data are neither acceptance corrected nor background subtracted, we follow the method described by LHCb~\cite{LHCb:2022fpg} to correct for the former. For the latter, we add a linearly rising, non-interfering background to the uncertainty. 

The fit is shown in Fig.~\ref{fig:DeltaK} as a red band; see Appendix~\ref{app:fit}~\cite{SuppMat} for details.
\nocite{LHCb:privcomm,Belle:2005rpz,BaBar:2008lpx}
Its stability within uncertainties has been tested against a gradual increase of the relative error for all data points, simulating a systematic error,
and against variations of the fit range below the two-kaon threshold.
We checked that including an isoscalar tensor partial wave (restricted to not exceeding the data in the $f_2(1270)$ mass range), or CPV real parts in the source constants,
does not significantly improve the data description.  Moreover, we checked that releasing the constraints on the $\omega$ couplings to the source and $b_{S0s}=0$ for ${\cal A}_{S0s}^{\pm}$ returns values consistent with the constraints themselves.
The error bands are generated using bootstrapping and imposing the confidence limits on the $\chi^2/\text{d.o.f.}$ as prescribed in Ref.~\cite{Press:2007ipz}. 

{\it Discussion}---The relevance of different contributions to the two CPV distributions is illustrated in Fig.~\ref{fig:ContributionsImportant}.
The advantage of the approach presented here is that its parameters have a clear physical meaning.  For example, 
the  CP-even imaginary parts of the source terms, $c_{i}$, are due to $c\bar c$ loops, related to open-charm meson pairs in hadronic terms.  
The presence of a $\rho$ peak in $\Delta \Gamma^{(+)}_\text{CP}$, as seen in  Figs.~\ref{fig:DeltaK} and \ref{fig:ContributionsImportant}, illustrates the need for such a contribution, since according to Eq.~\eqref{eq:mainresult} it scales as $c_{P}b_{P}\vert \Omega_P\vert^2$. 
Moreover, the isotensor $S$~wave, ${\cal A}_{S2}^\pm$, is essential for the description of $\Delta \Gamma^{(+)}_\text{CP}$, interfering with the strange and non-strange isoscalar $S$~waves. It is also required for an accurate description of $\Delta \Gamma^{(-)}$, interfering with the $\rho$.
The isotensor amplitude was not considered in earlier studies, since it does not contain any resonance.

\begin{figure}
\includegraphics[width=\linewidth]{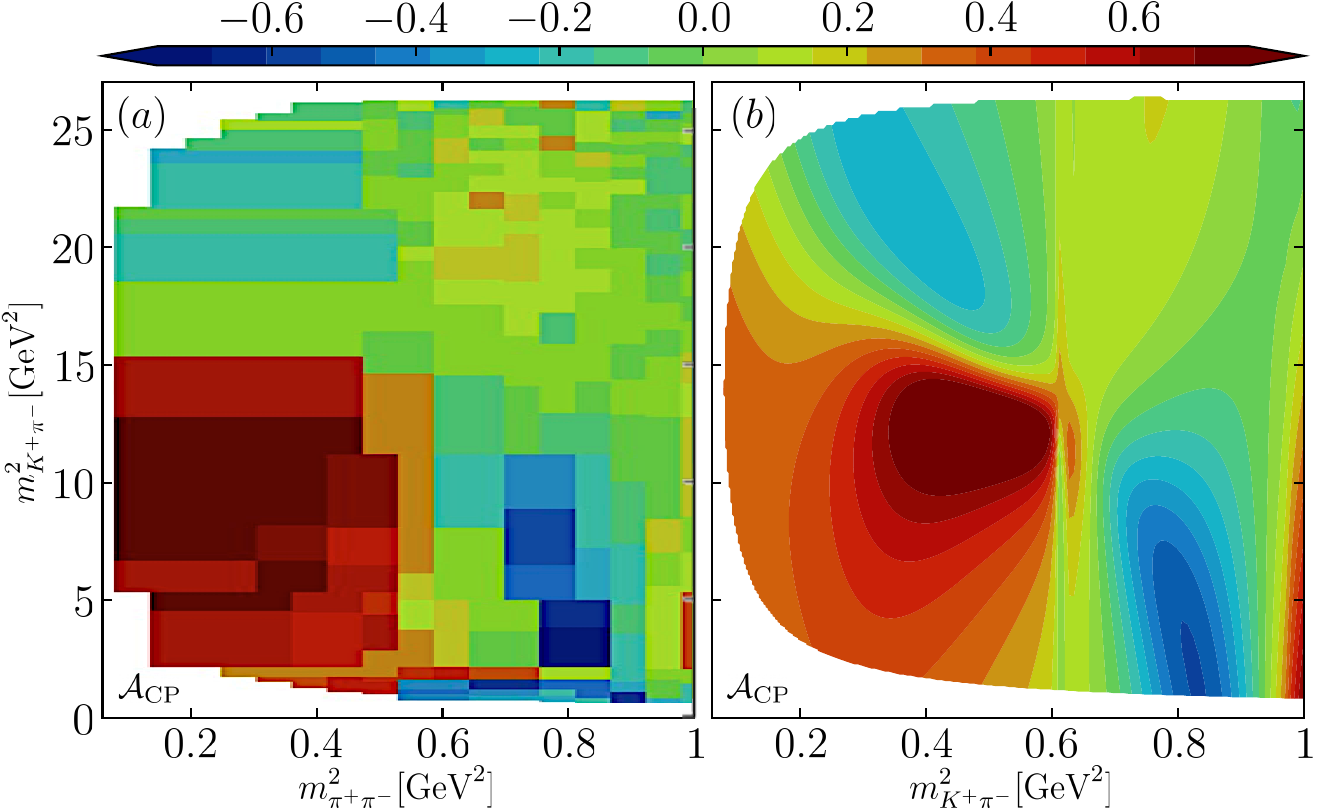}
\caption{
$B^\pm\to K^\pm\pi^+\pi^-$ CP asymmetry, distributed over the Dalitz plot section $m_{\pi^+\pi^-}^2\leq 1\GeV^2$.
(a) LHCb binned raw asymmetry ${\cal A}_\text{CP}$, cropped and enlarged from Fig.~3 in Ref.~\cite{LHCb:2022fpg}.  
(b) ${\cal A}_\text{CP}$ from our analysis. 
}
\label{fig:Dalitz}
\end{figure}

Finally, we have only fitted projected CP-odd distributions. However, our primary goal is to describe the CPV asymmetry in $B^\pm\to K^\pm \pi^+\pi^-$ observed by LHCb in localized regions of their Dalitz plot, which reads
\begin{equation}
    {\cal A}_\text{CP}(s,t)=\frac{\Delta\Gamma_\text{CP}(s,z(s,t))}{\Sigma\Gamma(s,z(z,t))}=\frac{\Gamma^-(s,t)-\Gamma^+(s,t)}{\Gamma^-(s,t)+\Gamma^+(s,t)}.
\end{equation}
Since $\Gamma^\pm(s,t)\geq0$, it follows that $\vert{\cal A}_\text{CP}(s,t)\vert\leq1$.
Given that our approach is limited to $s\leq1\GeV^2$, in Fig.~\ref{fig:Dalitz} we only show the corresponding low-$s$ section of the Dalitz plot. 
In Fig.~\ref{fig:Dalitz}(a), we have cropped and enlarged our region of interest for the LHCb raw asymmetry from Fig.~3 in Ref.~\cite{LHCb:2022fpg}. Note the use of an adaptive binning with a constant number of events per bin.

With all parameters fixed, our model predicts the ${\cal A}_\text{CP}$ Dalitz plot, which
is shown in Fig.~\ref{fig:Dalitz}(b). The agreement between the data and our model is remarkable, considering that both the data and our results correspond to central values whose uncertainties cannot be shown in this plot.
In particular, we reproduce the two localized regions with large CPV, $\vert{\cal A}_\text{CP}(s,t)\vert \geq60\%$, shown in dark red or blue.
Since a small CP-even denominator amplifies ${\cal A}_\text{CP}$, the predicted CP-violating numerator $\Delta\Gamma_\text{CP}$ 
and CP-even denominator are provided in Appendix~\ref{app:Dalitz}~\cite{SuppMat}.

{\it Summary and conclusions}---We have presented a dispersive method to quantify the contribution of low-invariant-mass pion--pion rescattering to final-state interactions in $B^\pm\to K^\pm\pi^+\pi^-$ and their CP-violating asymmetries. The advantage of our method is
that it can incorporate high-precision universal $\pi\pi$ low-energy interactions. This also includes the commonly neglected non-resonant isospin-2 contribution,  which plays an essential role. Once the parameters are fixed from the angular-integrated decay data from LHCb, we can predict the CP-asymmetry Dalitz-plot distribution, including the remarkable presence of large localized CP asymmetries. Our approach illustrates the role of low-energy final-state interactions in enhancing CP violation in localized kinematic regions. 
It could be applied to other regions of the Dalitz plot and to any
$B^\pm\to h_1^\pm h_2^+ h_2^-$ decays, for $h_n^\pm=\pi^\pm,K^\pm$. 
Our findings will be even more relevant since LHCb will soon collect four times more data in Run 3 and forty times more in HL-LHC~\cite{LHCb:2018roe}.

\bigskip

{\it Acknowledgments}---This work was formulated and initiated during the three-month research visit of JRP to Bonn University under the MCIN ``Salvador de Madariaga'' Program, Grant No.\ PRX22/00129, as well as during the preparations of the newly accepted Color meets Flavor
cluster, funded by the Deutsche Forschungsgemeinschaft
(DFG, German Research Foundation) under Germany’s
Excellence Strategy---EXC 3107---Project-ID 533766364.
Financial support by the MKW NRW under funding code No.\ NW21-024-A, 
by the Konrad-Adenauer-Stiftung e.V.\ with funds from the
BMBF, 
the Spanish Grant No.\ PID2022-136510NB-C31 funded by MCIN/AEI/10.13039/501100011033, 
the European Union's Horizon 2020 research and innovation program under Grant Agreement No.\ 824093 (STRONG2020), 
the Brazilian funding FAEPEX Grant No.\ 2030-24, and INCT-FNA CnPq grant,
is gratefully acknowledged.
T.M.\ was supported by the Deutsche Forschungs\-gemeinschaft (DFG, German Research Foundation) under Grant No.\ 396021762 -- TRR 257 ``Particle Physics Phenomenology after the Higgs Discovery.''
C.H.\ thanks the CAS President's International Fellowship Initiative (PIFI) under Grant No.\ 2025PD0087 for partial support. 

\appendix

\section*{Supplemental Material}
\label{AppendixA}

\section{Fit details}
\label{app:fit}

We provide here further details of our fitting procedure and best fit result. We have extracted the data and their uncertainties from the figures in Ref.~\cite{LHCb:2022fpg}.
 Our best fit corresponds to a  $\chi^2/\text{d.o.f.}=3.4$.
However, in addition to possible concerns about the accuracy of this procedure to obtain the data, we believe this $\chi^2/\text{d.o.f.}$ has no strict statistical interpretation, because
the projected data are neither corrected for acceptance effects (efficiency, production asymmetry, \ldots) nor background subtracted. 
Since the acceptance correction is only provided for the integrated yields,  we have corrected all bins with that very same average correction. Moreover, we also corrected for background with a very naive estimate $b(s)\propto b (\sqrt s-2M_\pi) $, trying to mimic the projections of the combinatorial background used by LHCb in their ongoing analysis~\cite{LHCb:privcomm}.

\begin{figure}[t]
    \centering
    \includegraphics[
  width=1\linewidth,
  trim=0 0.3cm 0 0.7cm,
  clip
]{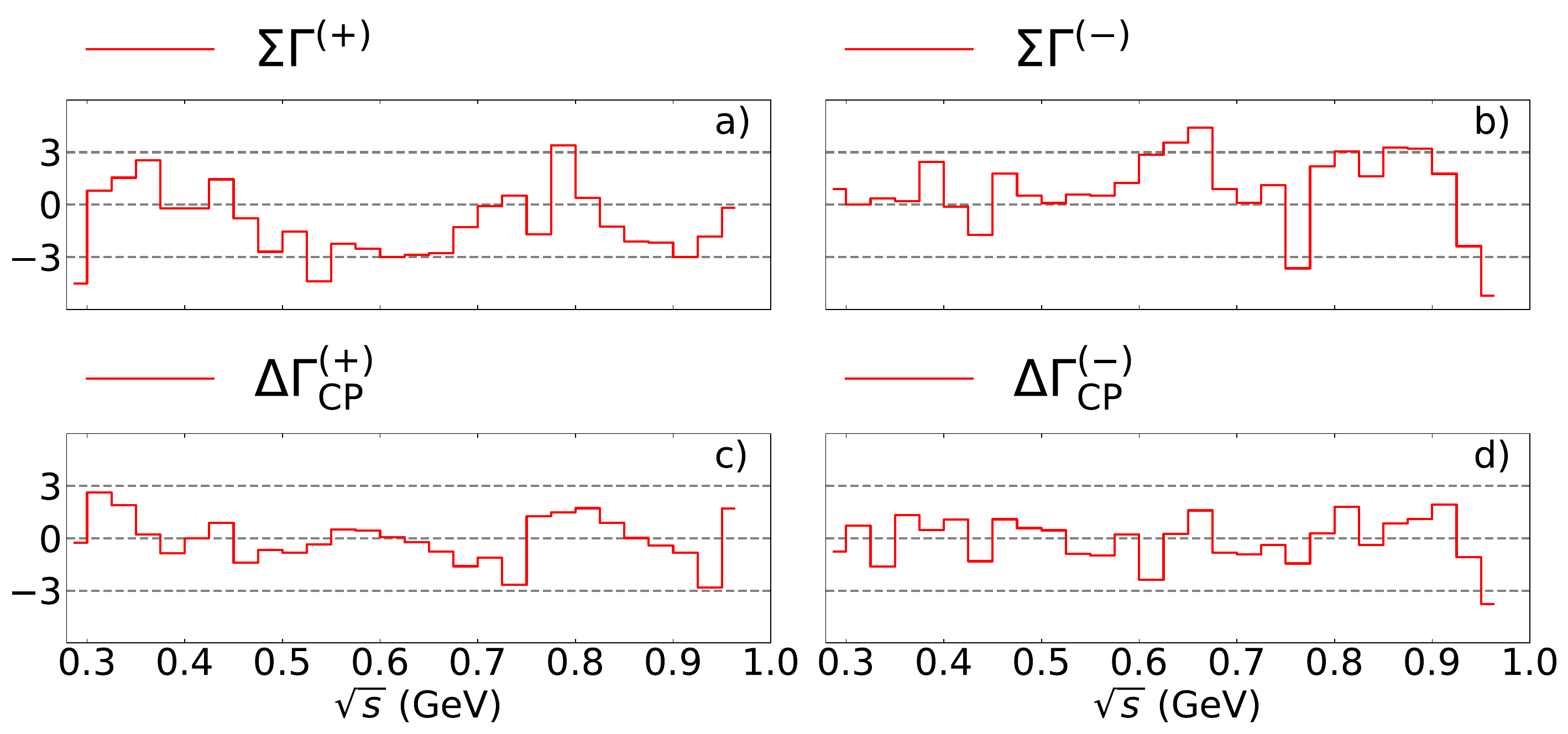}
    \caption{Pulls of our fit, corresponding to the four panels in Fig.~\ref{fig:DeltaK}. 
    }
    \label{fig:Pulls}
\end{figure}

Note that not even the LHCb collaboration itself provides $\chi^2/\text{d.o.f.}$ values, neither in their Dalitz-plot amplitude analyses  nor for their projections (see, e.g., that of $B\to3\pi$ in Ref.~\cite{Aaij:2019jaq}).
Furthermore, none of the previous theoretical analyses of Belle~\cite{Belle:2005rpz} and BaBar~\cite{BaBar:2008lpx} data, like those of Refs.~\cite{Simma:1991ct,Fajfer:2004cx,Furman:2005xp,El-Bennich:2006rcn,El-Bennich:2009gqk}, which fit projections directly, provide the corresponding $\chi^2/\text{d.o.f.}$ values.

To compare our fit quality with that obtained by LHCb in their projections~\cite{Aaij:2019jaq}, in Fig.~\ref{fig:Pulls} we provide the pulls of our fit. They are presented in the same order as the fits in Fig.~\ref{fig:DeltaK}.
These are very similar to the ones shown in, e.g., Fig.~6 of the LHCb amplitude analyses in Ref.~\cite{Aaij:2019jaq}.

In addition, we have checked that a naive $\chi^2/\text{d.o.f.}\sim1$ can be obtained by adding an uncorrelated $\sim7\%$ systematic uncertainty to all yields.
This provides an estimate of the size of systematic uncertainties and the effect of the approximations in our approach. The fit parameters remain within their uncertainties under this artificial increase in errors. 

Let us recall that our aim here is to provide a sound formalism, not to perform a proper statistical analysis of these data, whose details we do not possess.  Such an analysis should be carried out with access to the necessary input, using our method.

\begin{figure}[h]
    \centering
    \includegraphics[width=1\linewidth]{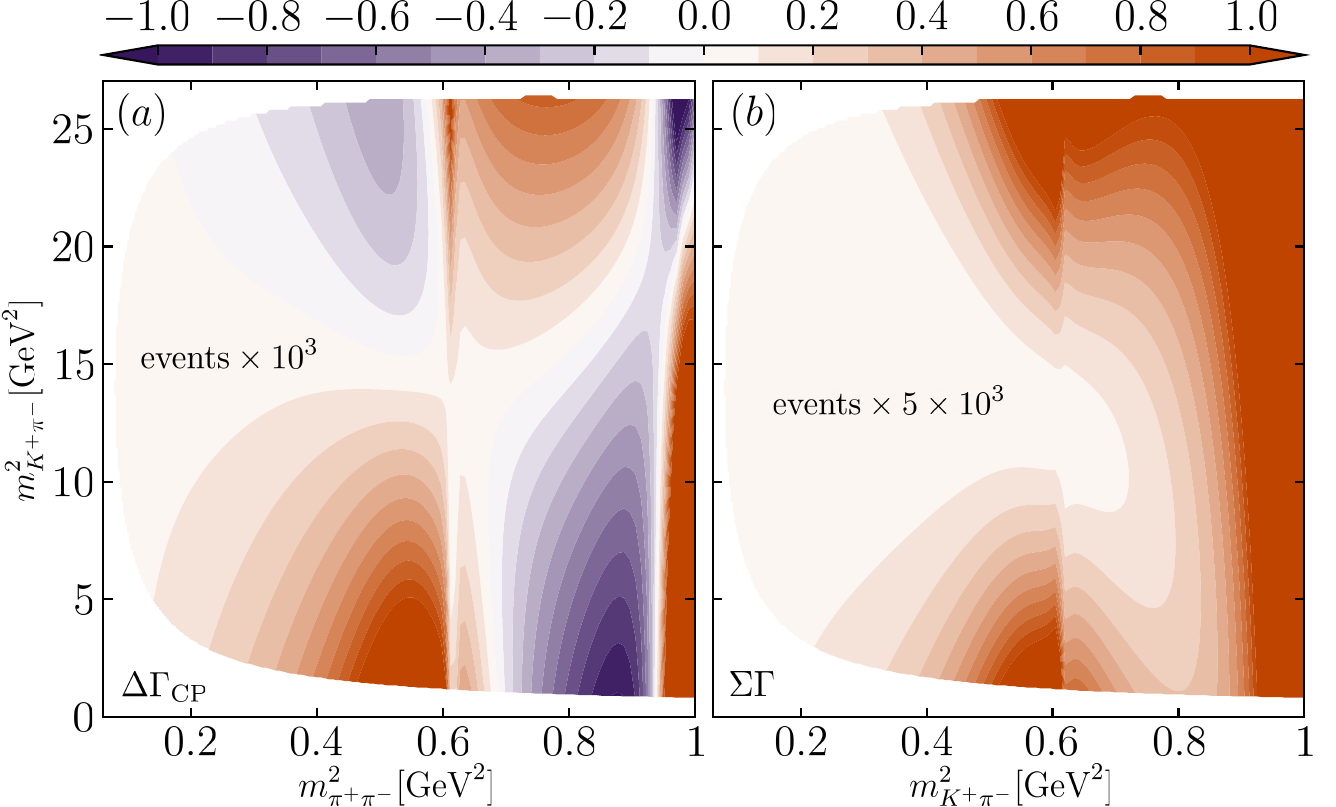}
    \caption{Numerator and denominator of the $B^\pm\to K^\pm\pi^+\pi^-$ CP asymmetry ${\cal A}_{\text CP}$, shown in Fig.~\ref{fig:Dalitz}, distributed over the Dalitz plot section $m_{\pi^+\pi^-}^2\leq 1\GeV^2$. 
Left: the CP-violating numerator $\Delta\Gamma_\text{CP}$. Right: the CP-conserving denominator $\Sigma\Gamma$. Note the different scales.\\}
    \label{fig:NumDen}
\end{figure}
\section{Dalitz-plot distribution }
\label{app:Dalitz}

In Fig.~\ref{fig:NumDen}, we show the numerator and denominator needed to obtain the ratio ${\cal A}_\text{CP}$, whose Dalitz-plot distribution is shown in Fig.~\ref{fig:Dalitz}.  The CP-violating numerator $\Delta\Gamma_\text{CP}$ is shown in the left panel and the CP-even denominator is shown in the right panel. 
By comparing with Fig.~\ref{fig:Dalitz}, note that a small CP-even denominator amplifies ${\cal A}_\text{CP}$.

\bibliography{CPV_refs}

\end{document}